\documentclass[twocolumn]{aastex63}

\usepackage{upgreek}
\usepackage{hhline}
\usepackage{amsmath}
\usepackage[LGRgreek]{mathastext}
\usepackage{multirow}
\shorttitle{UCD Radio obs.}
 \shortauthors{Hughes et al.}

\begin{document}

\title{Unlocking the Origins of Ultracool Dwarf Radio Emission}

\author{A. G. Hughes}
\affiliation{Department of Physics and Astronomy,
University of British Columbia,
6224 Agricultural Rd.,
Vancouver, BC V6T 1Z1, Canada}

\author{A. C. Boley} 
\affiliation{Department of Physics and Astronomy,
University of British Columbia,
6224 Agricultural Rd.,
Vancouver, BC V6T 1Z1, Canada}

\author{R. A. Osten} 
\affiliation{Space Telescope Science Institute,
 3700 San Martin Drive,
 Baltimore, MD 21218, USA }
\affiliation{Center for Astrophysical Sciences, 
Johns Hopkins University,
Baltimore, MD 21218, USA}

\author[0000-0001-8445-0444]{J. A. White}
\affiliation{National Radio Astronomy Observatory, 520 Edgemont Rd., Charlottesville, VA, 22903, USA}
\affiliation{Jansky Fellow of the National Radio Astronomy Observatory}

\author{M. Leacock} 
\affiliation{Department of Physics and Astronomy,
University of British Columbia,
6224 Agricultural Rd.,
Vancouver, BC V6T 1Z1, Canada}



\begin{abstract}


Empirical trends in stellar X-ray and radio luminosities suggest that low mass ultracool dwarfs (UCDs) should not produce significant radio emission. Defying these expectations, strong non-thermal emission has been observed in a few UCDs in the $1{-}10$ GHz range, with a variable component often attributed to global aurorae and a steady component attributed to other processes such as gyrosynchrotron emission. While both auroral and gyrosynchrotron emission peak near the critical frequency, only the latter radiation is expected to extend into millimeter wavelengths.  We present
ALMA 97.5 GHz and VLA 33 GHz observations of a small survey of 5 UCDs. LP 349-25, LSR
J1835+3259, and NLTT 33370 were detected at 97.5 GHz, while LP 423-31 and LP 415-20 resulted in non-detections at 33 GHz. A significant flare was observed in NLTT 33370 that reached a peak flux of $4880 \pm 360\,  \mu Jy$, exceeding the quiescent flux by nearly an order of magnitude, and lasting 20 seconds. These ALMA observations show bright 97.5 GHz emission with spectral indices ranging from $\alpha = –0.76$ to $\alpha = –0.29$, suggestive of optically thin gyrosynchrotron emission. If such emission traces magnetic reconnection events, then this could have consequences for both UCD magnetic models and the atmospheric stability of planets in orbit around them.
Overall, our results provide confirmation that gyrosynchrotron radiation in radio loud UCDs can remain detectable into the millimeter regime. 

\end{abstract}
\section{Introduction}
\indent
Ultracool dwarfs (UCDs) are brown dwarfs and low-mass M stars of spectral type later than M7.  M dwarfs are the most common stellar objects in the Milky Way \citep{henry2006}, yet because they are so faint, UCDs are difficult to study.  They are also not expected to behave in the same way as more massive M-type stars, which are notorious for their flaring and variability \citep{davenport}.  Unlike more massive stars, UCDs are fully convective and lack a tachocline, or shearing layer between convective and radiative regions \citep{Chabrier2000}. Since the tachocline is thought to play a crucial role in the solar dynamo, it was long assumed that UCDs would be unable to generate solar strength magnetic fields and corresponding radio emission \citep{mohanty2002}. 

Observations of magnetically active F through early-M stars show a clear trend known as the G{\"u}del-Benz relation (GBR),  a tight correlation between the X-ray and radio luminosities \citep{gbr}.  The relation is described well by a single power law and extends through 10 orders of magnitude, suggesting a common emission mechanism. If UCDs were to follow the GBR, most would have no detectable radio emission whatsoever.  This makes the detection of some UCDs at  radio frequencies particularly mysterious. Highly sensitive observatories such as the NSF's Karl G. Jansky Very Large Array (VLA) and Arecibo have revealed that about 10\% of UCDs are bright at radio frequencies, with $1{-}10$ GHz emission exceeding the GBR by up to four orders of magnitude \citep{berger2006,Antonova2013,Route2016,Lynch2016}.  Strong magnetic fields up to kG strength have been measured in UCDs \citep{Reiners_2006, Reiners2009, Reiners_2010}, and inferred from radio emission \citep[e.g.][]{Kao2018}.  The overall radio detection rate of $10\%$ curiously increases to nearly $50 \%$ for objects with $\textit{v}\,sin\textit{i} > 20~km\,s^{-1}$.  The reason for this switch is unclear and the $50 \%$ detection rate is determined from a sample of ${<}\,20$ targets \citep{mclean2012}.

UCDs occupy the mass range in the transition between giant planets and stars, and observations indicate that radio active UCDs are capable of behaving like either one \citep{Berger2001, berger2006, Hallinan2006, mclean2012, Williams_2015}. Auroral emission prompted by the electron cyclotron maser instability (ECMI)  \citep{Williams2014, Hallinan2015, Kao2016, pineda2017} is akin to what is observed in all Solar System planets possessing a magnetic field \citep{zarka2001}, whereas synchrotron and gyrosynchrotron radiation are responsible for incoherent millimeter emission from the Sun and other stars  \citep{dulk1985, gudel2002, Bastian1998, Osten2008, osten2009}.

Bright, pulsed emission is more characteristic of aurorae, although \citet{Hallinan2006} suggest that auroral emission could become depolarized during propagation through the magnetosphere. In contrast, quasi-quiescent emission (i.e. steady emission with small variations in magnitude) is largely consistent with gyrosynchrotron radiation \citep{Berger2002, berger2006}. Early radio detections of UCDs were in the $1-10 \,\, GHz$ range, where both emission mechanisms are expected to peak for most UCD magnetic field strengths. These first detections often showed both a quiescent component, attributed to synchrotron or gyrosynchrotron emission, and a bursting component, usually attributed to ECMI \citep{liebert, burg_put, berger2006,  hallinan, WilliamsBerger}.  It was unclear from these detections alone whether the presumed gyrosynchrotron component would extend to higher radio frequencies. However, the first detection of a UCD by 
\citet{Williams_2015} at millimeter wavelengths confirms the presence of optically thin gyrosynchrotron radiation and measures the spectral index of the emission in this regime.

While auroral emission and gyrosynchrotron radiation can both be present in the same object, gyrosynchrotron radiation in the optically thin regime can most effectively be characterized using observations at high radio frequencies. In particular, gyrosynchrotron radiation will dominate between $20-100$ GHz for typical UCD magnetic field strengths, while ECMI emission becomes unlikely for frequencies above the fundamental frequency \citep{MelroseDulk1982, Treumann2006}. For lower magnetic field strengths, this frequency range will be shifted to lower frequencies as well. The first detection of the non-accreting UCD TVLM 513-46546 in this range could consequently not be attributed to ECMI, leaving optically thin gyrosynchrotron radiation as the presumed emission mechanism \citep{Williams_2015}.

Each emission mechanism traces different underlying magnetic processes and field configurations. The presence of auroral emission could suggest a global largely dipolar magnetic field \citep{Kao2016, pineda2017, berdyugina2017}, in which the co-rotation breakdown between magnetosphere plasma and the UCD magnetic field enables electrons to drift along electromagnetic currents \citep{Cowley2001}. Gyrosynchrotron emission, per \citet{Williams2014}, could be produced by reconnection events resulting from a non-axisymmetric field topology at higher-order field components. Both situations can occur simultaneously, with a global dipolar magnetic field producing auroral emission and localized reconnection events producing gyrosynchrotron radiation, potentially explaining the apparent concurrence of both mechanisms in radio bright UCDs.

UCDs are likely to have a high occurrence rate of terrestrial planets \citep{dresschar2015}. This presents a different possibility; planets in orbit around UCDs may be capable of inducing auroral radio emission in an analogous way to Io around Jupiter \citep{Turnpenney2018}.  However, should gyrosynchrotron emission and the corresponding reconnection events also be present, then the stability of planetary atmospheres could be threatened by the outgoing energetic particle flux \citep{segura, tilley}. 

In this paper, we present high radio frequency observations ($30-100$ GHz) of five UCDs. Prior to this work, the only non-accreting UCD to be detected in this frequency range was TVLM 513-46546, which \citet{Williams_2015} found to have 95 GHz emission consistent with gyrosynchrotron radiation. We find that two of our five targets are radio quiet over the course of the observations, and place upper flux limits on their emission. The detected UCDs represent the second through fourth non-accreting UCDs found to emit at millimeter wavelengths.

\begin{table*}

\centering 
\begin{tabular}{c | c  c  c  c  c} 
\hline\hline \ 
   	Property & LP 349-25 AB  & LSR J1835$+$3259 & NLTT 33370 & LP 415-20 & LP 423-31  \\
   		
   	\hline 
   	\multirow{2}{*}{Spectral Type} & \multirow{2}{*}{ M8$^{[1]}$} & \multirow{2}{*}{ M8.5$^{[5]}$ } & \multirow{2}{*}{M7$^{[7]}$} & \multirow{2}{*}{M9.5$^{[10]}$} & \multirow{2}{*}{ M7$^{11}$}\\
     & & & & & \\
    
    \hline 
    
   	\multirow{2}{*}{$M_{tot}$ ($M_{\odot}$)} & \multirow{2}{*}{ $0.120^{+0.008[2]}_{-0.007}$ } & \multirow{2}{*}{ $0.05 \pm 0.0038^{[6]}$ }   & \multirow{2}{*}{  $0.179^{+0.055[8]}_{ -0.062}$ } &  \multirow{2}{*}{ $0.09 \pm 0.06$ } &   \multirow{2}{*}{ $0.01 \pm 0.003^{[12]}$ }  \\
     & & & & & \\

   	\hline 
   	
   \multirow{2}{*}{	Distance (pc) } & \multirow{2}{*}{ 10.3$\pm 1.70^{[3]}$ } & \multirow{2}{*}{ $5.67 \pm 0.02^{[5]}$ }   & \multirow{2}{*}{ $16.39 \pm 0.75^{[7]}$  } &  \multirow{2}{*}{ $21 \pm 5^{[10]}$ } &   \multirow{2}{*}{ $10.7 \pm 0.3^{[12]}$}  \\
   	 & & & & & \\
   
   \hline 
   
   	\multirow{2}{*}{Age (Myr) }& \multirow{2}{*}{ - } & \multirow{2}{*}{ $22\pm 4^{[6]}$ }   & \multirow{2}{*}{  $\sim30-200^{[8]}$ } &  \multirow{2}{*}{ - } &   \multirow{2}{*}{- }  \\
   	& & & & & \\
   
   \hline 
   
   	\multirow{2}{*}{ $T_{eff}$ (K) }& \multirow{2}{*}{ - } & \multirow{2}{*}{ $2800 \pm 30^{[6]}$ }   &   $3200 \pm 500^{[8]}$  & $2600 \pm 100^{[10]}$ &   \multirow{2}{*}{ -}  \\
   	 & & & $3100 \pm 500^{[8]}$& $2400 \pm 100^{[10]}$ & \\
   	
   	\hline 
   	
   	\multirow{2}{*}{$\textit{v} \, sin \textit{i}$ $(km ~ s^{-1})$} & $55 \pm 2^{[4]}$  & \multirow{2}{*}{ $50 \pm 5^{[6]}$ }   & \multirow{2}{*}{  $45 \pm 5^{[9]}$ } &  $44 \pm 4^{[4]}$ &   \multirow{2}{*}{ $9.0 \pm 2^{[11]}$}  \\
   	 &  $83 \pm 3^{[4]}$& & & $36 \pm 4^{[4]}$& \\
   	
   	\hline 
   	
   	\multirow{2}{*}{Obs Frequency (GHz)} & \multirow{2}{*}{ 97.5 } & \multirow{2}{*}{ 97.5 }   & \multirow{2}{*}{  97.5 } &  \multirow{2}{*}{ 97.5 } &   \multirow{2}{*}{ 33.0 }  \\
   	 & & & & & \\
   	
   	\hline 
   	
   	\multirow{2}{*}{Flux (mJy)} & \multirow{2}{*}{ $70\pm15$ } & \multirow{2}{*}{ $114\pm21$ }   & \multirow{2}{*}{ $604\pm61$  } &  \multirow{2}{*}{ $<20$ } &   \multirow{2}{*}{ $<30$}  \\
   	 & & & & & \\

\hline	

\end{tabular}
\caption{Properties of ultracool dwarfs presented in this work, including combined spectral type, total mass ($M_{tot}$), stellar distance, stellar age, effective temperature ($T_{eff}$), $\textit{v} \, sin \textit{i}$, frequency of the observations (GHz), and observed flux (mJy). \\\hspace{\textwidth} References: [1] \citet{gizis} [2] \citet{Konopacky2010} [3] \citet{schmidt} [4] \citet{konopacky2012} [5] \citet{reid2003} [6] \citet{berdyugina2017} [7] \citet{Lepine2009} [8] \citet{Schlieder} [9] \citet{mclean2011} [10] \citet{Siegler2003} [11] \citet{rein_bas} [12] \citet{cruz_2003}. }
 \label{TBL:lims}
\end{table*}

\section{Observations and Results}
In this section, we discuss each of the five targets used in this study along with the corresponding observations. The results and target list are reported in Table \ref{TBL:lims}.

\begin{figure*}
\centering
 \includegraphics[width=\linewidth]{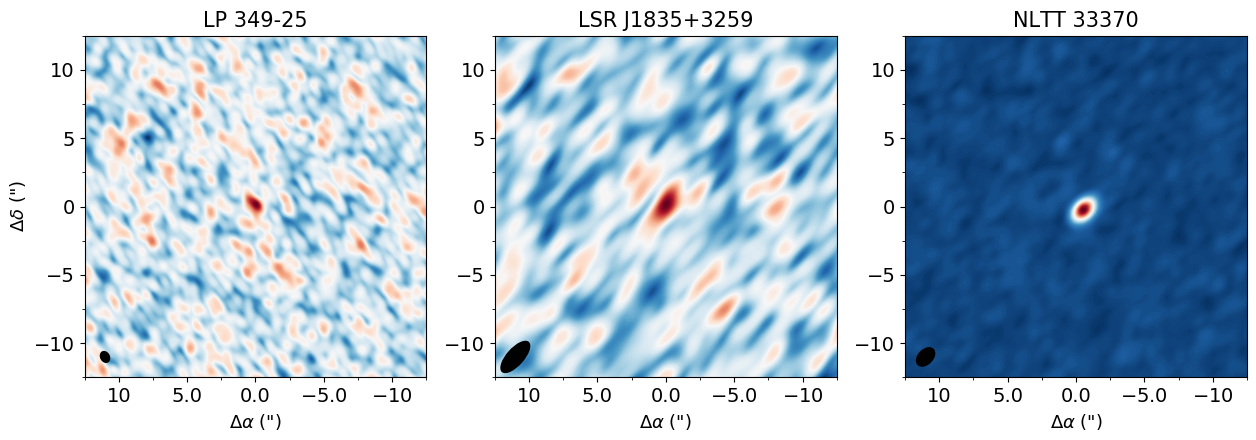}
\caption{All detected UCDs. The visibilities are phase-shifted to account for the proper motion. LP 349-25 has peak flux of 70 $\mu Jy$ with RMS of 9 $\mu Jy$, LSR J1835+3259 has a peak flux of 114 $\mu Jy$ with RMS 18 $\mu Jy$, and NLTT 33370 has a peak quiescent flux of 604 $\mu Jy$ with RMS 7.3 $\mu Jy$. The synthesized beam is denoted by the black ellipse in the bottom left of each image.}
\label{fig:test}
\end{figure*}

LP 349-25, LSR J1835$+$3259, and NLTT 33370  are detected and exhibit bright radio emission at 97.5 GHz. All three of these targets were also detected at lower radio frequencies in previous studies \citep[e.g.,][]{phanbao2007, hallinan2008, mclean2011}. The images of the detected sources are given in Figure \ref{fig:test}. In contrast, the observations of LP 423-31 and LP 415-20 yield null detections at 97.5 GHz and 33 GHz, respectively. LP 423-31 had been previously observed at 8.46 GHz, which also resulted in a null detection, while there are no previous radio observations of LP 415-20. For both targets, we use the non-detections to place upper limits on the radio fluxes.

\subsection{LP 349-25}

LP 349-25 is a spectroscopic binary of spectral types M8 and M9, with rapid rotation speeds of $55 \pm 2$ and $83 \pm 3 ~\rm km ~s^{-1}$, respectively \citep{konopacky2012}. It was previously detected at 8.46 GHz by \citet{phanbao2007} at a flux density of $365 \pm 16 ~ \mu Jy$, at $8.3$ and $5$ GHz at flux densities of $365 \pm 16 ~\mu Jy$ and $329 \pm 38 ~\mu Jy$, respecively by \citet{osten2009} and by \citet{mclean2012} at $323 \pm 14 \, \mu Jy$, also at 8.46 GHz.

The archival Atacama Large \\ Millimetre/Submillimeter Array (ALMA) Cycle 4 observations \newline (ID 2016.1.00817.S, PI Williams) were centered on J2000 coordinates RA = 00h 27m 55.99s $\delta$ = +22d 19' 32.80". The data were taken in 3 execution blocks (EBs) from 2016 October 22 to 2016 November 6 
for a total of 2 hr on-source (2.9 hr including overhead). There were 45 antennas used with baselines ranging from 15 m to 500 m. 

The Band 3 observations have a total bandwidth of 8 GHz split among 4 spectral windows (SPW). Each SPW has $128\times 15.625$ MHz channels. The SPWs are centred at 90.4 GHz, 92.4 GHz, 102.5 GHz, and 104.5 GHz, giving an effective continuum frequency of 97.5 GHz. The data were reduced using {\scriptsize CASA 5.4.1} \citep{casa_ref}, which included water vapor radiometer (WVR) calibration; system temperature corrections; flux and bandpass calibration with quasar J0423-0120; and phase calibration with quasar J0431$+$1731. The average precipitable water vapor (PWV) was 2.08 mm throughout the observations.

We generated CLEANed images using the {\scriptsize CASA} task \textit{tclean}, natural weighting down to a threshold of the RMS noise, and a cell size of $\frac{1}{10}$ the beam size.
These ALMA 97.5 GHz observations achieve an RMS sensitivity of $9~ \upmu \textrm{Jy}~\textrm{beam}^{-1}$ as measured from the CLEANed image away from the source.
The size of the resulting synthesized beam is $0\farcs82  \times 0\farcs62$ at a position angle of $28.3^{ \circ }$, corresponding to $7$ au at the system distance of $10.3 \pm 1.70$ pc \citep{schmidt}.  We use the {\scriptsize CASA} task \textit{uvmodelfit} to fit a point source to the visibilities, we find a best fit flux of $70 \pm 9 \, \mu Jy$. The data were subsequently split by observing scan, and both the flux density and RMS of each scan were taken using \textit{uvmodelfit} to generate a time series. No flaring activity was seen over the course of the observations.

\subsection{LSR J1835$+$3259}

LSR J1835$+$3259 is a brown dwarf of spectral type M8.5 \citep{reid2003} with a rapid rotation speed of $\textit{v} \, sin \textit{i} =50 \pm 5 \rm ~km ~s^{-1}$ known to be variable across multiple wavelengths. It has been detected at 8.44 GHz with both quiescent, largely unpolarized emission and $100 \%$ circularly polarized bursting emission, with an average flux density of $722 \pm 15 \, \mu Jy$ and variable emission reaching up to $2500 \, \mu Jy$ \citep{hallinan2008}.  

The archival ALMA Cycle 4 observations \\ (ID 2016.1.00817.S, PI Williams were centered on J2000 coordinates RA = 18h 35m 37.88s $\delta$ = +32d 59' 53.31". The data were taken in 2 EBs from 2016 November 6 to 2016 December 1 for a total of 1 hr on-source (2 hr including overhead). There were 40 antennas used with baselines ranging from 15 m to 460 m.

The Band 3 observations have a total bandwidth of 8 GHz split among 4 SPWs. Each SPW has $128\times 15.625$ MHz channels. The SPWs are centered at 91.5 GHz, 93.4 GHz, 101.5 GHz, and 103.5 GHz, giving an effective continuum frequency of 97.5 GHz. The data were reduced using {\scriptsize CASA 4.7.2}, which included WVR calibration; system temperature corrections; flux and bandpass calibration with quasar J2025$+$3343; and phase calibration with quasar 3 J1848$+$3219. The average PWV was 2.53 mm throughout the observations.

We generated CLEANed images using the {\scriptsize CASA} task \textit{tclean}, natural weighting down to a threshold of the RMS noise, and a cell size of $\frac{1}{10}$ the beam size.
These ALMA 97.5 GHz observations achieve an RMS sensitivity of $18~ \upmu \textrm{Jy}~\textrm{beam}^{-1}$ as measured from the CLEANed image away from the source. The size of the resulting synthesized beam is $2\farcs85  \times 1\farcs18$ at a position angle of $-41.8^{ \circ }$, corresponding to $12$ au at the system distance of $5.6875 \pm 0.0005 \, pc$ \citep{gaiadr2}.  We use the {\scriptsize CASA} task \textit{uvmodelfit} to fit a point source to the visibilities, we find a best-fit flux of $114 \pm 18 \, \mu Jy$. The data were subsequently split by observing scan, and both the flux density and RMS of each scan were taken using \textit{uvmodelfit} to generate a time series. No flaring activity was seen over the course of the observations.

\subsection{NLTT 33370}
NLTT 33370AB (originally LPSM J1314+1320) is a resolved binary of combined spectral type M7.0e \citep{Law_2006, Lepine2009, Forbrich2016}.  This magnetically active system has been well-studied relative to other UCDs due to its brightness. Although the system is binary, we refer to it as NLTT 33370 for ease of reading.

NLTT 33370  has been detected with short, polarized bursts and quiescent, largely unpolarized radio emission, both of which are brighter than that observed in any other UCD. Quasi-quiescent emission was detected with the VLA at 8.46 GHz with a flux density of $1156 \pm 15 \,\, \mu Jy$ and at 22.5 GHz with a flux density of $763 \pm 84 \,\, \mu Jy$ \citep{mclean2011}. It was subsequently detected in a multi-wavelength study by \citet{Williams_NLTT}, exhibiting both quasi-periodic steady emission and flaring radio emission. These studies found that the steady emission was circularly polarized and modulated sinusoidally with a period of $3.89 \pm 0.05 \, hr$. \citet{Williams_NLTT} further found an additional unpolarized component in one of their two VLA monitoring campaigns. The observations are interpreted to include a combination of flaring due to the ECMI and stable emission due to gyrosynchrotron radiation.

The archival ALMA Cycle 2 observations \\ (ID 2013.1.00976.S, (PI Williams) were centered on J2000 coordinates RA = 13h 14m 20.38s $\delta$ = +13d 18' 58.34". The data were taken in 4 EBs from 2015 June 10 to 2015 June 12 for a total of 2.7 hr on-source (4.8 hr including overhead). There were 38 antennas used with baselines ranging from 25 m to 820 m.

The Band 3 observations have a total bandwidth of 8 GHz split among 4 SPWs, each with $64\times 31.25$ MHz channels. The SPWs are centred at 90.5 GHz, 92.4 GHz, 102.5 GHz, and 104.5 GHz, giving an effective continuum frequency of 97.5 GHz. The data were reduced using {\scriptsize CASA 4.7.0}, which included WVR calibration; system temperature corrections; flux and bandpass calibration with quasar J1229$+$0203; and phase calibration with quasar J1347$+$1217. The average PWV was 0.71 mm throughout the observations.

\begin{figure*}
\centering
 \includegraphics[width=1\linewidth]{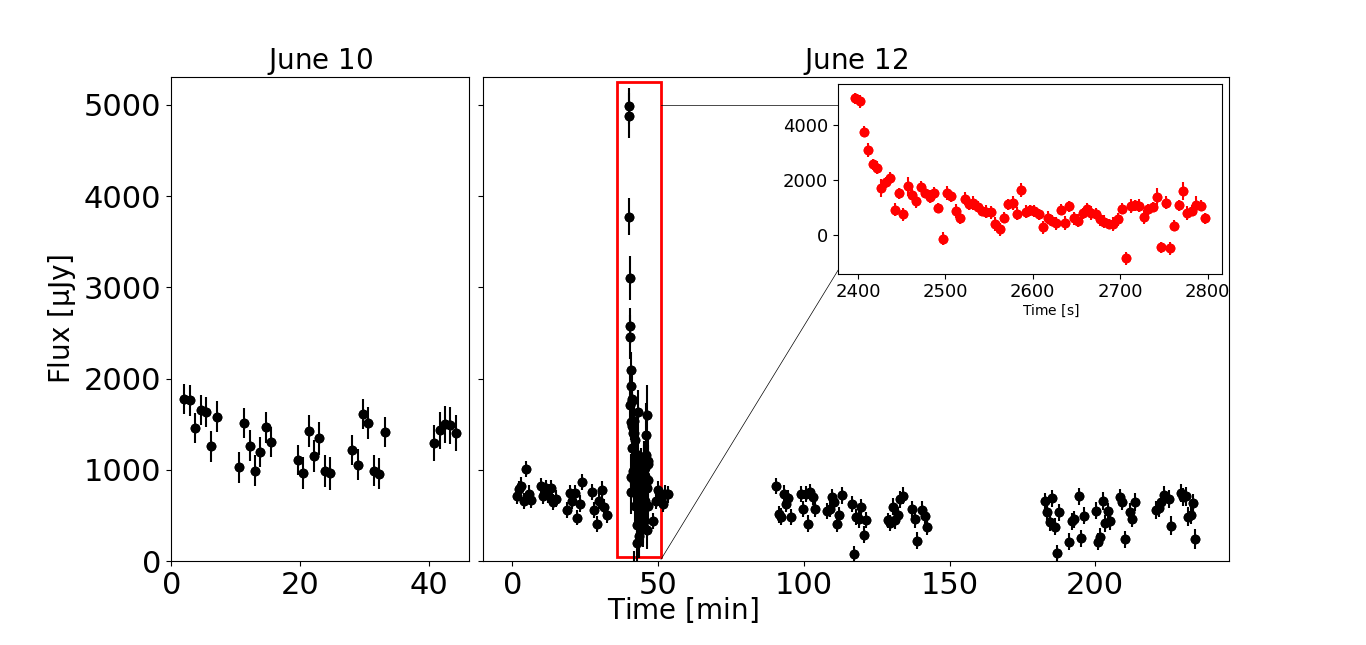}
\caption{Time series of all NLTT 33370 observations presented in this work, including time on June 10th and June 12th, 2015. The data set was split into 30-second bins, and the peak flux and RMS were taken using the {\scriptsize CASA} task \textit{uvmodelfit}. The two panels show the light curves of each day, while the overlayed plot shows the June 12 flare that reached a peak flux of $4880 \pm 360 ~\mu Jy$. Due to the gaps in the data, it is unclear whether this flare repeats with the stellar rotation.}
\label{fig:emitting}
\end{figure*}

We generated CLEANed images using the {\scriptsize CASA} task \textit{tclean}, natural weighting down to a threshold of the RMS noise, and a cell size of $\frac{1}{10}$ the beam size.
These ALMA 97.5 GHz observations achieve an RMS sensitivity of $9~ \upmu \textrm{Jy}~\textrm{beam}^{-1}$ as measured from the CLEANed image away from the source. The size of the resulting synthesized beam is $1\farcs58  \times 1\farcs04$ at a position angle of $-44.8^{ \circ }$, corresponding to $22$ au at the system distance of $16.39 \pm 0.75$ pc \citep{Lepine2009}. A 10-second flare (Fig. \ref{fig:emitting}) was also captured during these observations, with a peak flux of $4880 \pm 360 \, \mu Jy$ from the CLEANed image of the flare only. We use the {\scriptsize CASA} task \textit{uvmodelfit} to fit a point source to the visibilities, we find a best-fit flux of $604 \pm 7.3 \, \mu Jy$ for the quiescent emission.

\subsection{LP 415-20AB}
LP 415-20AB is a resolved binary with spectral types M7(A) and M9.5(B) and high $\textit{v} \, sin \textit{i}$ of $40 \pm 5$ and $37 \pm 4 \rm ~km \, ~s^{-1}$, respectively \citep{konopacky2012}. The presence of two UCDs with rapid rotation suggests that this system should be a radio emitter; however, we see no evidence of emission throughout the duration of our observations. 

The ALMA Cycle 6 observations (ID 2018.1.01088.S, PI Hughes) presented here were taken in 3 EBs from 2019 January 8 to 2019 January 21 for a total of 2 hr on-source (2.7 hr including overhead). There were 58 antennas used with baselines ranging from 15 m to 500 m. 

These Band 3 observations have a total bandwidth of 8 GHz split among 4 SPWs, each with $128\times 15.625$ MHz channels. The SPWs are centred at 90.5 GHz, 92.4 GHz, 102.5 GHz, and 104.5 GHz, giving an effective continuum frequency of 97.50 GHz. The data were reduced using {\scriptsize CASA 5.4.1}, which included WVR calibration; system temperature corrections; flux and bandpass calibration with quasar J0423-0120; and phase calibration with quasar J0431$+$1731. The PWV ranged from 2.9 mm to 5.9 mm throughout the observations.

We generated CLEANed images using the {\scriptsize CASA} task \textit{tclean}, natural weighting down to a threshold of the RMS noise, and a cell size of $\frac{1}{10}$ the beam size.
These ALMA 97.5 GHz observations achieve an RMS sensitivity of $6.7~ \upmu \textrm{Jy}~\textrm{beam}^{-1}$ as measured from the CLEANed image away from the source. The size of the resulting synthesized beam is $3\farcs26  \times 2\farcs91$ at a position angle of $30.1^{ \circ }$, corresponding to $65$ au at the system distance of 21 pc.  We report a 97.5 GHz non-detection of LP 415-20AB with a $3 {\sigma}$ upper level flux limit of 20 $\mu Jy$ measured from the CLEANed image.

\subsection{LP 423-31}
LP 423-31 is an M7 star with a relatively low rotation rate of  $9 \,\, km ~s^{-1}$ and strong magnetic field of $3500^{+400}_{-600}$ G \citep{rein_bas, mclean2011}. \citet{berger2006} placed an upper flux limit of $< 39 ~ \mu Jy$ using VLA observations at 8.46 GHz. This star is expected to be radio silent given the existing trends and models in UCD radio emission, but confirmation is needed.

The observations were taken during the VLA Semester 18B (ID VLA-18B-241, PI Hughes) over 2 scheduling blocks (SBs) on 2018 December 12 for a total of 12.8 minutes on-source (35.5 min including overhead). Data were acquired with the array in the C antenna configuration, with 26 antennas and baselines ranging from 35 m to 3400 m. 

The instrument configuration uses the Ka band receiver with a correlator setup consisting of $7808 \times 1$ MHz channels for a total bandwidth of 7.8 GHz. Four separate basebands are included with rest frequency centres at 30 GHz, 32 GHz, 34 GHz, and 36 GHz giving an effective continuum frequency of 33.0 GHz. The quasar J0750$+$1231 was used for gain and phase calibration and quasar 3C286 was used as a bandpass and flux calibrator. Data were reduced using the {\scriptsize CASA 4.7.2} pipeline, which included bandpass, flux, and phase calibrations. 

We generated CLEANed images using the {\scriptsize CASA} task \textit{tclean}, natural weighting down to a threshold of the RMS noise, and a cell size of $\frac{1}{10}$ the beam size.
These VLA 33 GHz observations achieve an RMS sensitivity of $10~ \upmu \textrm{Jy}~\textrm{beam}^{-1}$ as measured from the CLEANed image away from the source. The size of the synthesized beam is $0\farcs86 \times 0\farcs72$ at a position angle of 40.7$^{ \circ }$. The beam size corresponds to $8$ au at the system's distance of $10.3 \pm 1.70$  pc \citep{konopacky2012}.  We report a 33 GHz non-detection of LP 423-31 with a $3 \sigma$ upper level flux limit of $30 ~ \mu Jy$ measured from the CLEANed image.

The lack of emission at high radio frequencies of a slowly rotating UCD observed to be radio-quiet at lower frequencies is consistent with emerging trends in UCD radio emission that indicate rapid rotators are more likely to be radio bright \citep{mclean2011, pineda2017}. Our results show that UCDs active at 1-8 GHz may be likely to be bright at millimetre wavelengths as well.

\section{Discussion}

The five UCDs presented in this work have a range of properties, including low/high  $\textit{v} \, sin \textit{i}$ value, binarity, and detection/non-detection at low radio frequencies, making them a good selection for probing trends in high radio frequency emission. LP 349-25, LSR J1835$+$3259, NLTT 33370, and LP 423-31 have all been observed at lower radio frequencies. LSR J1835$+$3259, LP 349-45 and NLTT 33370 have been observed multiple times in the $1-10$ GHz range. In all cases the flux densities were consistent between observations. LP 423-31, however, has been detected in some 8.46 GHz observations (Stelzer et al. in prep) and not detected in others \citep{berger2006}. We find that all three UCDs that are consistently active at low radio frequencies are also active at high radio frequencies, where ECMI emission is cut off. Furthermore, if the emission is largely due to ECMI, then there could be some level of circular polarization present in the ALMA data. In principle, this would provide an additional metric to disentangle the emission mechanisms. Unfortunately, these data were taken in earlier ALMA cycles when polarization calibration was unreliable so no conclusive determination on the presence of polarization during the observations can be made. However, the inferred spectral indices, discussed below, and strength of emission at these high frequencies suggest that optically thin gyrosynchrotron radiation is active in all the detected UCDs. A larger sample of UCDs must be observed in this range to determine whether this small sample is representative of radio emitting UCDs as a whole.

\begin{figure}
\centering
 \includegraphics[width=1.\linewidth]{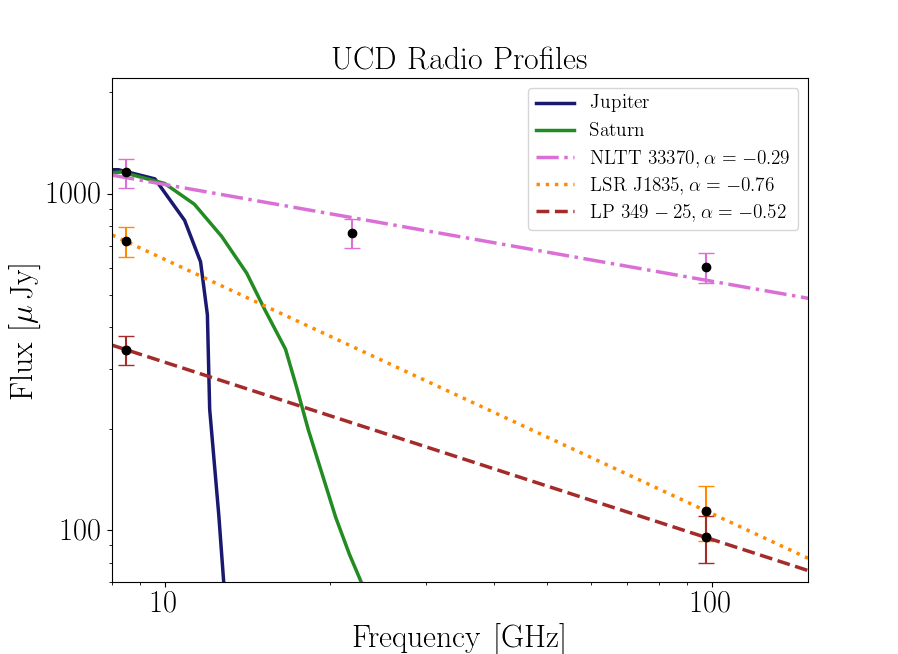}
\caption{Radio emission profiles of all three detected UCDs are shown. The kilometric radiation profiles of Jupiter and Saturn adapted from \citet{zarka} are plotted in blue and green, scaled and shifted to the peak flux and frequency of NLTT 33370, whereas the dotted lines correspond to fitted gyrosynchrotron models. The spectral indices of LP 349-25 and LSR J1835-46546 were determined by connecting the two existing data points for each UCD, and were $\alpha = -0.52$ and $\alpha = -0.76$, respectively.} 
\label{fig:rad_profiles}
\end{figure}

The measurements at different frequencies can be used to estimate a spectral index $\alpha$ such that $F_{\nu}\,\propto\,\nu^{\alpha}$ for flux density $F_{\nu}$ (Fig \ref{fig:rad_profiles}). Moreover, if gyrosynchrotron radiation is responsible for the detections, then the spectral indices can be related to an electron energy distribution following \citet{dulk_marsh}.  LSR J1835-46546 and LP 349-25 have spectral indices of $\alpha = -0.76 \pm 0.07$ and $\alpha = -0.52 \pm 0.06$, respectively, which are consistent with optically thin gyrosynchrotron radiation. NLTT 33370 is best fit by a spectral index of $\alpha = -0.29 \pm 0.04$, slightly more shallow than what would be expected for gyrosynchrotron or synchrotron radiation. Using the relation between $\alpha$ and $\delta$, we calculate a common electron index of $\delta \approx 2$, with individual values ranging from $2.20 \pm 0.08$ to $1.68 \pm 0.05$. Also plotted in Figure \ref{fig:rad_profiles} are ECMI profiles of Jupiter and Saturn, shifted to the peak flux and frequency of NLTT 33370, which illustrate the sharp cutoff of ECMI with increasing frequency compared with gyrosynchrotron radiation.

The three detected UCDs share two key characteristics: reliably detectable emission at 8.46 GHz and rotation speeds far in excess of the ad hoc $40 ~ \rm km~s^{-1}$ cutoff, where radio emission becomes much more likely \citep{mclean2011, pineda2017}. Previous detections of radio emission from UCDs have predominantly been in the $1-10$ GHz frequency range, where both ECMI and gyrosynchrotron emission are generally present. The brightness temperatures implied by the flux densities of observations presented in this work range from $\sim 10^6 - 10^8$ K. Incoherent processes are limited to brightness temperatures below approximately $10^{12}$ K \citep{Readhead1994}, whereas coherent processes such as ECMI can reach much higher brightness temperatures typically of order $10^{12}$ K and above \citep{Melrose1991}. The inferred brightness temperatures indicate that the detected millimeter emission is likely produced by gyrosynchrotron radiation. The emission from LSR J1835+3259 and LP 349-25 AB was quiescent over the respective 1-hour and 2-hour observations, with no evidence of flaring or variability. While the emission of NLTT 33370 was mostly quasi-quiescent, the June 12 observations included a massive 20-second burst, with emission over 5 times brighter than the quiescent flux.

The origin of the NLTT 33370 burst is unclear, but consistent with previous observations of the system at lower frequencies. NLTT 33370 was found by \citet{Williams_NLTT} to have both bursts and quiescent flux modulated with two distinct periodicities of $3.7859 \pm 0.0001 $ and $3.7130 \pm 0.0002$ hours, slightly less than the stellar rotation period of 3.9 hours \citep{mclean2011}. The flare presented in this work did not recur with either of these periods.  ECMI bursts due to auroral emission are expected to occur with the stellar rotation period.  Unfortunately, none of the on-source observing blocks corresponded to a 3.9 hour interval from the time of the burst.  Given the presence of quiescent emission at 97.5 GHz, the burst could  be associated with gyrosynchrotron radiation released during a major flare.  Long-term monitoring of NLTT 33370 at high radio frequencies is necessary to determine if such emission is quasi-periodic or stochastic.

The other UCDs in the sample, LP 415-20 and LP 423-31 were not detected.  We place $3 \sigma$ upper flux density limits of 21 $\mu Jy$ and 30 $\mu Jy$, respectively. LP 415-20 has not been previously observed at low radio frequencies, but its high $\textit{v} \, sin \textit{i}$ makes  the non-detection surprising compared with the other targets and is an important target for follow up.  While we did not detect emission at 33 GHz, the slowly rotating LP 423-31 has been observed to have variable and highly polarized emission at 8 GHz (Stelzer et al. in prep). The lack of 33 GHz emission, in conjunction with the prior 8 GHz detection, might suggest that ECMI is present in this object and quiescent gyrosynchrotron is weak, not present at all, or does not extend to millimeter wavelengths. However, because previous observations by \citet{berger2006} at 8.46 GHz have also reported non-detections, additional observations of this target are required to confirm whether it is also variable at high radio frequencies and possibly inactive over the course of the 35 minute observations presented in this work.

The measured flux can be used to estimate the size of the emitting region. Assuming the low frequency limit of $h \nu << k_B T_b$, which is valid for the frequency of observation, the Rayleigh-Jeans law can be rearranged to give, 
\begin{equation}
    X_{R_J} \approx 1400 \frac{d_{\rm pc}}{\nu_{\rm GHz}} \sqrt{\frac{S_{\rm \mu Jy}}{T_b}},
 \label{eqn:emitrgn}
\end{equation}
where $d_{\rm pc}$ is the stellar distance, $\nu_{\rm GHz}$ is the observing frequency, $S_{\rm \mu Jy}$ is the flux density, $T_b$ is the brightness temperature, and $X_{R_J}$ is the size of the emitting region in units of Jupiter radii.  In the case of an isolated source with constant non-thermal effective temperature, the brightness temperature can be related to the non-thermal effective temperature by $T_b = (1 - e^{- \tau(\nu)}) T_{\textrm{eff}}$, where $\tau(\nu)$ is the optical depth.  This gives the size of the emitting region in the optically thin regime,
\begin{equation}
    X_{R_J} \approx 1400  \frac{d_{\rm pc}}{\nu_{\rm GHz}} \sqrt{\frac{S_{\mu Jy}}{(1 - \exp^{- \tau(\nu)}) T_{\textrm{eff}}}}.
\label{eqn:opt_thin_xrj}
\end{equation}
The precise value of the optical depth for gyrosynchrotron emission depends critically on the angle between the magnetic field, the line-of-sight to the observer \citep{Zheleznyakov}, and the frequency of observations.

\begin{figure}
 \centering
 \includegraphics[width=1.\linewidth]{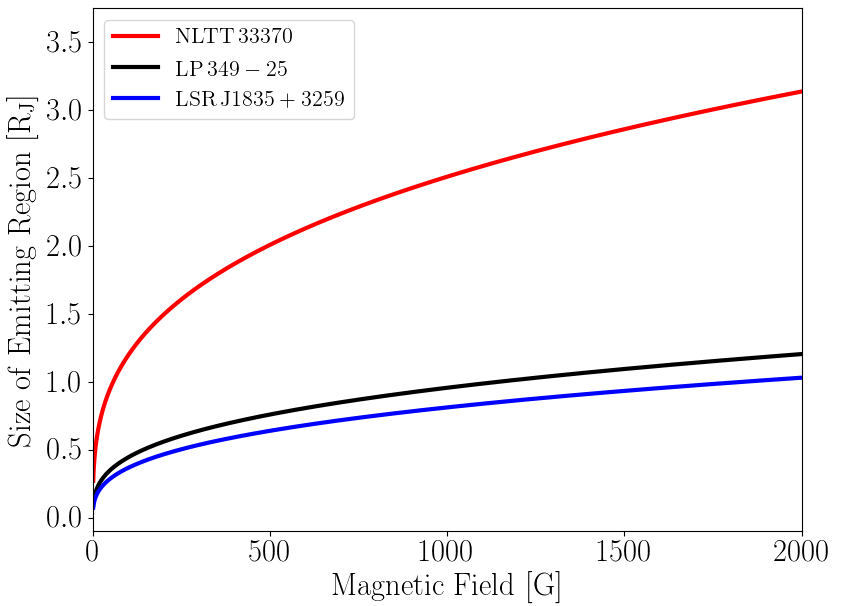}
\caption{The size of the emitting region for a given magnetic field strength for all detected UCDs.  The solid curves show the size of the emitting region given $\sim 8.5$ GHz flux density of each UCD, assuming this is in the transition region where $\tau \approx 1$.  The size of the emitting region is given in Jupiter radii $R_{\rm J}$, characteristic of UCD sizes, while the magnetic field range is chosen to be representative of Solar to typical UCD values.}
\label{fig:curves}
\end{figure}

The size of the emitting region is plotted against magnetic field strength for all three UCDs in Figure \ref{fig:curves}. The solid lines are determined from the previously published ${\sim} 8.5$ GHz flux densities of all UCDs. The emitting region is calculated by following Equation \ref{eqn:emitrgn}, where we are assuming the observations are at approximately the peak emission frequency where $\tau \approx 1$ and $T_b$ approaches $2/3 T_{\textrm{eff}}$ following the non-thermal \citet{dulk_marsh} relations. Taking the average UCD radius to be that of Jupiter \citep{sorahana_2013}, these results show that LP 349-25 and LSR J1835$+$3259 have emitting regions ranging between the full stellar disk at large magnetic field strengths (${\sim} 2000$ G) to a fraction of the stellar disk for low magnetic field strengths. If the 97.5 GHz flux densities originate from the same emitting region, then we expect an optical depth of $\tau \approx 0.0001$.

The values spanned by this range are consistent with observations of M stars that show significant portions of the disk covered with magnetically active regions \citep{johnskrull, alekseev}, showing a magnetic filling factor - the fraction of surface covered by magnetically active area - significantly higher than Solar values (which are typically ${<} 1 \%$). The emitting region for NLTT 33370 is larger than the stellar disk for magnetic field strengths $>100$ G. This enhanced emitting region is possibly owing in part to the binarity of the system. If both NLTT 33370 A and NLTT 33370 B are radio loud, then the inferred size of the emitting region includes contributions from both stellar disks and accounts for an additional factor of up to $\sqrt{2}$. However, this cannot explain the predicted emitting region of NLTT 33370 for magnetic field strengths $>$170 G. If the magnetic field strength exceeds this value, the size of the emitting region could be enhanced by the effect of a companion.

As detailed by \citet{Williams2014}, gyrosynchrotron radiation could originate in frequent small-scale magnetic reconnection events. These events accelerate a population of downward-directed electrons to near-relativistic energies, prompting detectable radio emission, while also injecting highly energetic particles into the stellar environment. The outgoing energetic particles cannot be detected directly, but gyrosynchrotron emission produced by the mildly relativistic electrons can be used as a tracer for such events \citep{Bastian1998, Osten2016, Hughes2019}. While several studies have shown that stellar UV emission alone may not be catastrophic for planets and could even be beneficial \citep{Ranjan2016}, simulations ran by \citet{segura} and \citet{tilley} suggest that it is energetic particles, not UV emission alone, that is the most damaging to planetary atmospheres. If gyrosynchrotron radiation is indeed produced by reconnection events and is commonplace around the 10\% of UCDs that are radio active, then its presence might be an important consideration for modeling of exoplanet atmospheres in orbit around such UCDs.

\section{Summary}

In this work we presented ALMA and VLA observations of five UCDs with a range of properties and radio behavior. Of these targets, three were found to be active at 97.5 GHz, whereas two were not detected at 33 GHz (LP 423-31) or 97.5 GHz (LP 415-20). Thus far, all UCDs that have been reliably detected near 8 GHz and within $30-100$ GHz exhibit emission in both of these frequency ranges.   The observations presented here give spectral indices ranging from $\alpha = -0.76$ to $\alpha = -0.29$, suggestive of optically thin gyrosynchrotron emission. 
The data for all detected UCDs were split into smaller time bins, with the peak flux and RMS taken from each chunk using the {\scriptsize CASA} task \textit{uvmodelfit}. Each peak flux was then used to generate a time series of the stellar brightness. Two of the detections (LP 349-25 and LSR J1835) show little time variation throughout the observations (2 hr and 1 hr on source, respectively), while NLTT 33370 shows minor temporal variability and a strong flare that exceeded the average quiescent flux by a factor of 20. 
Of the non-detections, LP 415-20 is a very rapid rotator that has not previously been observed at radio frequencies. Our observations at 33 GHz show it to be radio quiet, consistent with most UCDs but at odds with other rapidly rotating UCDs. LP 423-31  exhibits no detectable high radio frequency emission during the observations. This particular source has been observed twice previously, one of which resulted in a detection (Stelzer et al. in prep) and the other in a non-detection \citep{berger2006} at low radio frequencies. The presence of variable emission at low radio frequencies and the lack of detectable quiescent emission at high radio frequencies is suggestive of ECMI. However, the low $\textit{v} \, sin \textit{i}$ of this source muddles this picture. Overall, these results provide further evidence for the presence of millimeter emission radio loud UCDs, although additional observations are necessary to determine how common such emission might be. If the gyrosynchrotron radiation is produced by reconnection events, then a high resulting energetic particle flux can threaten the stability of planetary atmospheres \citep[e.g.][]{Hughes2019}. 

This work is part of an ongoing project to characterize the high radio frequency emission of UCDs. As of yet, very few observations of UCDs have been made in the $20 - 100 \, GHz$ range where the emission mechanism could be more readily distinguished through the emission's spectral index. With additional observations of a larger sample, we will gain a better understanding of UCD radio behavior.

\section*{Acknowledgements}
We thank the referee for the valuable feedback that improved the manuscript. This work was supported in part by an NSERC Discovery Grant, The University of British Columbia, The Canadian Foundation for Innovation, the BC Knowledge Development Fund and the Canada Research Chairs program.  This paper makes use of the following ALMA data: ADS/JAO.ALMA\#2016.1.00817.S, ADS/JAO.ALMA \#2013.1.00976.S, ADS/JAO.ALMA\#018.1.01088.S. ALMA is a partnership of ESO (representing its member states), NSF (USA) and NINS (Japan), together with NRC (Canada), MOST and ASIAA (Taiwan), and KASI (Republic of Korea), in cooperation with the Republic of Chile. The Joint ALMA Observatory is operated by ESO, AUI/NRAO and NAOJ. The National Radio Astronomy Observatory is a facility of the National Science Foundation operated under cooperative agreement by Associated Universities, Inc.

\bibliographystyle{aasjournal}



\end{document}